\documentclass[a4paper, 10pt, conference]{ieeeconf}      

\IEEEoverridecommandlockouts                              

\usepackage{graphicx} 
\usepackage{mathptmx} 
\usepackage{amsmath} 
\usepackage{amssymb}  
\usepackage{tikz}
\usepackage{pgfplots}
\usepackage{multirow} 
\usepackage{listings} 
\usepackage{subcaption} 
\usetikzlibrary{calc}   

\newcommand\copyrighttext{%
	\footnotesize \copyright{ }2020 IEEE. Personal use of this material is permitted. Permission from IEEE must be obtained for all other uses, in any current or future media, including reprinting/republishing this material for advertising or promotional purposes, creating new collective works, for resale or redistribution to servers or lists, or reuse of any copyrighted component of this work in other works.}
\newcommand\copyrightnotice{%
	\begin{tikzpicture}[remember picture,overlay]
	\node[anchor=south,yshift=10pt,xshift=7pt] at (current page.south) {\parbox{\dimexpr\textwidth-\fboxsep-\fboxrule\relax}{\copyrighttext}};
	\end{tikzpicture}%
}

\title{\LARGE \bf
Generation of Complex Road Networks Using a Simplified Logical Description for the Validation of Automated Vehicles*
}

\author{Daniel Becker$^{1}$, Fabian Ru{\ss}$^{1}$, Christian Geller$^2$ and Lutz Eckstein$^{3}$
\thanks{*This research is funded by the SET Level 4to5 research initiative, promoted by the	Federal Ministry for Economic Affairs and Energy (BMWi).}
\thanks{$^{1}$Daniel Becker and Fabian Ru{\ss} are with the automated driving department of the Institute for Automotive Engineering (ika), RWTH Aachen University, 52074 	Aachen, Germany {\tt\small \{daniel.becker, fabian.russ\}@ika.rwth-aachen.de}}%
\thanks{$^{2}$Christian Geller is with RWTH Aachen University, 52062 Aachen, Germany
	{\tt\small christian.geller@rwth-aachen.de}}%
\thanks{$^{3}$Lutz Eckstein is head of the Institute for Automotive Engineering (ika), RWTH Aachen University, 52074 Aachen, Germany {\tt\small lutz.eckstein@ika.rwth-aachen.de}}%
}

\begin{document}

\maketitle
\thispagestyle{empty}
\pagestyle{empty}
\copyrightnotice

\begin{abstract} 
Simulation is a valuable building block for the verification and validation of automated driving functions (ADF). When simulating urban driving scenarios, simulation maps are one important component. Often, the generation of those road networks is a time consuming and manual effort. Furthermore, typically many variations of a distinct junction or road section are demanded to ensure that an ADF can be validated in the process of releasing those functions to the public. 

Therefore, in this paper, we present a prototypical solution for a logical road network description which is easy to maintain and modify. The concept aims to be non-redundant so that changes of distinct quantities do not affect other places in the code and thus the variation of maps is straightforward. In addition, the simple definition of junctions is a focus of the work. Intersecting roads are defined separately, are then set in relation and the junction is finally generated automatically. 

The idea is to derive the description from a commonly used, standardized format for simulation maps in order to generate this format from the introduced logical description. Consequently, we developed a command-line tool that generates the standardized simulation map format OpenDRIVE.

\end{abstract}

\section{Introduction}
In recent years, the interest in autonomous driving has considerably increased, both in research and society. To ensure the safety of automated driving functions (ADF), verification and validation methods need to be developed and established. The approval of an ADF for public road traffic demands compliance with safety requirements regarding its functionality. The requirements include an error-free behavior over several billion test kilometers. For example Wachenfeld and Winner \cite{wachenfeld2016release} state that the distance to prove an interstate pilot amounts to 6.62 billion test kilometers, with a theoretically 50\% chance this prove would say that the ADF is twice as good as a human driver. Since this distance seems unreal and economically not acceptable, Schuldt \cite{schuldt2017beitrag} advise a scenario based-testing. In order to shorten the test time, save costs and detect functional errors at an early development stage, simulation is well suited. It is also appropriate to create any number of scenarios and to explore critical driving situations without endangering people.

Scenario-based testing introduces three different stages to describe scenarios systematically: functional, logical and concrete, cf.~\cite{menzel2018scenarios}. The term functional states a linguistic description which allows experts to talk about scenarios at an early stage of the development process. In logical scenarios, parameters and ideally their probability distributions are provided. Finally, concrete scenarios specify a distinct value for each parameter which makes them feasible to be executed reproducible in a simulation or on proving grounds. 

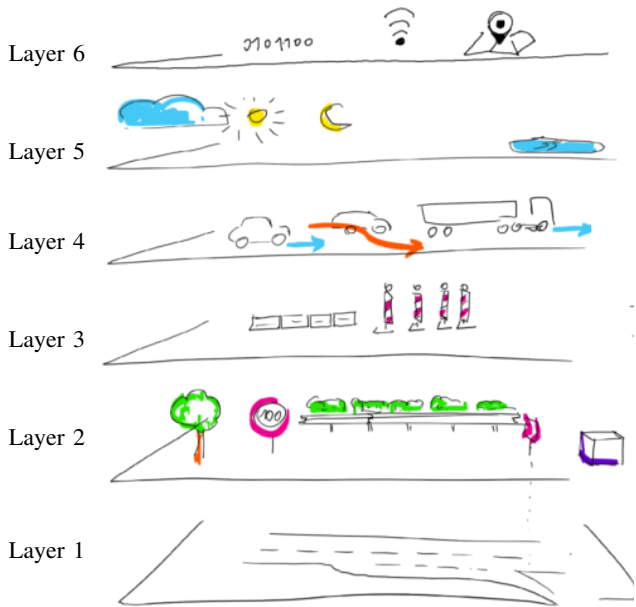
\begin{figure}[t]
	\begin{tikzpicture}
	\node[anchor=west](layer6) at (0,7.4) {\small Layer 6};
	\node[anchor=west](layer5) at (0,6.1) {\small Layer 5};
	\node[anchor=west](layer4) at (0,4.9) {\small Layer 4};
	\node[anchor=west](layer3) at (0,3.6) {\small Layer 3};
	\node[anchor=west](layer2) at (0,2.3) {\small Layer 2};
	\node[anchor=west](layer1) at (0,0.8) {\small Layer 1};
	\begin{scope}[xshift=1.22cm]
	\node[anchor=south west,inner sep=0] (image) at (0,0) {\includegraphics[width=0.4\textwidth]{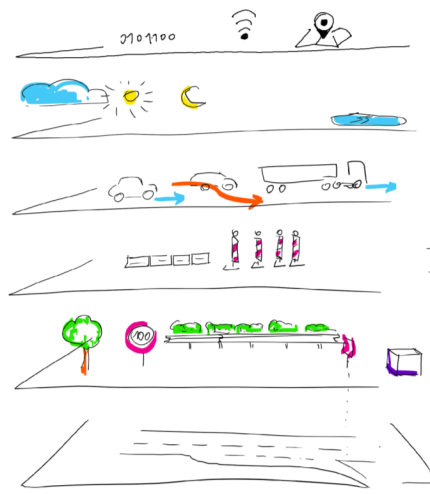}};
	\end{scope}
	\end{tikzpicture}
	\caption{The data layers for the description of a scenario following~\cite{bock2018data} and~\cite{bagschik2018ontology}.}
	\label{fig_layers}
\end{figure}

This kind of description uses the subdivisions of the six-layer model described by Bock et al. in \cite{bock2018data}, which is based on the layer model in Schuldt \cite{schuldt2017beitrag}. In this model, a scenario is divided in basic components and only an interaction between all the six layers represents a complete scenario. The model is illustrated in Fig.~\ref{fig_layers}. The three lower layers describe the static part of the scenario. Layer four focuses on moving objects, whereas layer five and six describe environmental conditions and vehicle-to-everything (V2X) communication, respectively. Detailed discussion of layer four on German motorways can be found in Weber et al. \cite{weber2019framework}.

Besides layer four which describes the dynamic behavior of all traffic participants, the static scenery has to be taken into account as well. The latter consists of the road network and all non moving objects which are relevant for the scenario. As mentioned, it is described within layer one to three of the six layer model and has to be defined in a logical and concrete manner, respectively. The open source standard format OpenDRIVE \cite{odr1.5} is suitable for concrete descriptions of road networks and is widely used in industry and research. However, by now there is no standardized logical road description format which is easy to use and capable of being transferred into OpenDRIVE.
\begin{figure}[t] 		
	\centering
	\includegraphics{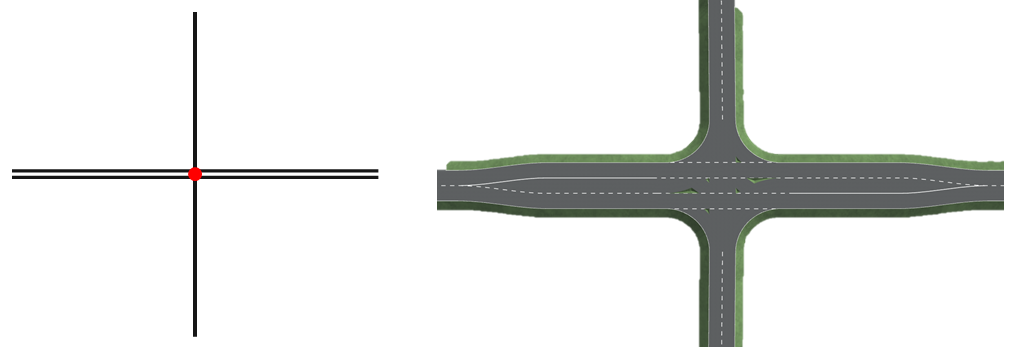}
	\caption[dummy]{Graphical example of input and output of the road generation tool. As a minimal input, two reference lines with assigned road types and a coupling point are sufficient to generate an intersection written in valid OpenDRIVE, visualized in CarMaker 8.0\footnotemark.}
	\label{fig_motivation}
\end{figure}
\footnotetext{https://ipg-automotive.com/de/produkte-services/simulation-software/}

Current research initiatives such as SET Level 4to5 which is funded by the Federal Ministry for Economic Affairs and Energy (BMWi) of Germany, address simulative scenario-based testing with a focus on inner city traffic. Consequently, the concept presented in this paper aims to offer a description for road networks which allows easy modifications of the map due to a non-redundant design. Further, it is possible to provide limited information for intersections and lane definitions since some default configurations and dependencies for main and access roads are defined. The input shall be translated into a concrete standardized road description format. Therefore, with the logical description comes a road generation tool that calculates all required OpenDRIVE quantities from the simplified input. We use XML to store the input and the corresponding OpenDRIVE output is saved in XML as well. This way we attempt to close a gap in the pipeline of defining logical scenarios on intersections and make them executable in a concrete manner.

An example which outlines the idea of the concept is shown in Fig.~\ref{fig_motivation} where an X-junction is defined by a main road intersected by an access road as the input and the result is a verified OpenDRIVE map which can be used in common automotive simulation software.

The remainder of this paper is organized as follows. First, Sec.~\ref{sec_relwork} introduces related work to the topic, followed by Sec.~\ref{sec_logicalDesc} where input format is described in more detail. Then, details on the implementation are documented in Sec.~\ref{sec_roadGen}. Next, in Sec.~\ref{sec_res} several results are presented and finally a conclusion and outlook is given in Sec.~\ref{sec_concl}.

\section{Related work} \label{sec_relwork}
During the research initiative PEGASUS \cite{pegasus.2019} the concepts of scenario based testing have been developed and utilized within the domain German motorway. Furthermore, the main focus has been on layer four. Motorways usually consist of parallel lanes which follow curves with big radii. Such a road is not complicated to describe and generate. It can be done automatically with low effort. Menzel et al. \cite{menzel2019functional} propose a method to create a standardized road network in the OpenDRIVE format from a logical (and functional) road description. 
A similar approach has been done by Noyer et al. \cite{dlrODRgen} who created a format called SimplifiedRoad which is an XML schema to describe long streets without junctions. This logical description is primary meant to easily create concrete highway sections and cannot specify all details in OpenDRIVE. However, when moving the testing domain to inner city scenarios such as the driving behavior on intersections, layers one to three (cf. Fig.~\ref{fig_layers}) are much more complex both in the logical description and in terms of the concrete format. 

There are basically two different ways of creating virtual roads for simulation in the industry. In the first variant, the streets are build based on recorded real data. Disadvantages are the effort of collecting sufficient information of road networks and the restricted possibility in track variation. Examples for such systems are Road2Simulation presented by Richter and Scholz in \cite{dlr110094} and services of the company atlatec\footnote{https://www.atlatec.de/}.
The second variant is based on the manual creation of a the track with a visual track editor. This way, a user is required to assemble the road manually. Examples of such systems are Road Designer \cite{ROD}, CarMaker (IPG Automotive) and others. Those tools are useful for creating distinct road networks in an easy to use way but an automated generation of a huge number of similar road networks is hard to realize.

\section{Logical road description}\label{sec_logicalDesc}
\begin{figure}[thpb] 		
	\centering
	\includegraphics{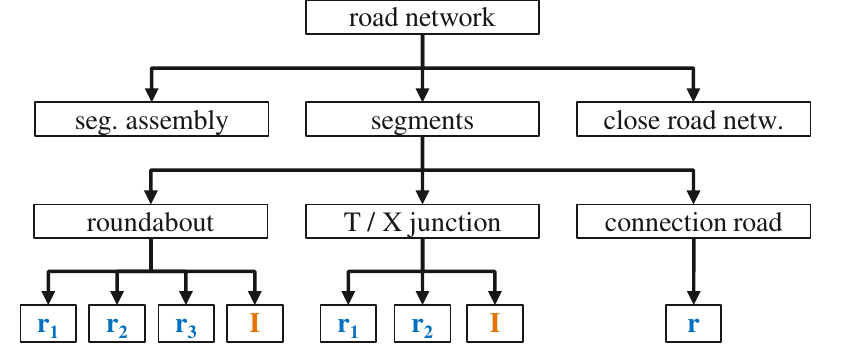}
	\caption{Hierarchical structure of the road network. On the first level all segments and their relation to each other are defined. In addition, it is possible to define the automatic connection of two road ends. Segments are defined as roundabouts, junctions and connection roads. $r_i$ and $I$ describe roads and intersection definitions, respectively.}
	\label{fig_schema}
\end{figure}

The proposed logical road description is derived from the reference line concept used in OpenDRIVE \cite{odr1.5}. This means that each lane is described by an offset from a reference line defined in the $x,y$-plane (topview). However, in contrast to OpenDRIVE we introduce some simplifications and eliminate redundancies along the reference line. Each road segment consists of the geometric primitives line, arc and spiral, which is motivated by the federal guidelines for road constructions in Germany cf. \cite{Baier.2008} and \cite{StraenNRW.2019}. A simplified illustration of the road network's XML structure is shown in Fig.~\ref{fig_schema}. On the first level under the \texttt{roadNetwork} tag all \texttt{segments} are defined and links which connect them can be stated (node "seg. assembly"). In addition, "loose ends" of segments can be automatically connected within the \texttt{closeRoadNetwork} tag. As illustrated, segments might be T-/X-junctions, roundabouts and connection roads. 

In the following sections, the reference line is examined in more detail, followed by the description of the input for the logical road network concept. The documentation is based on the nodes shown in Fig.~\ref{fig_schema}. 
\subsection{Reference Lines} \label{sec_refline} 
\begin{figure}
	\centering
	\includegraphics{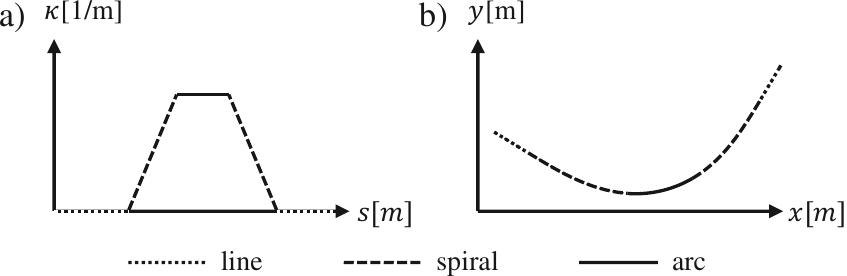}
	\caption{Qualitative example for a curve defined by the curvature graph. The plan view results from two integrations over $s$. Geometric primitives: line-spiral-arc-spiral-line. Positive curvature values correspond to left turns in positive $s$ direction.}
	\label{fig_curvGraph}
\end{figure}

In order to ensure a smooth driveability along the road network, we define that the curvature has to be continuous along roads where curvature is defined as the reciprocal of the radius. Therefore, the definition of curvature over distance, a start point and initial heading is sufficient to uniquely describe a road's course. At the same time, this implies that in contrast to OpenDRIVE it is not necessary to state the start point and heading of each geometric element (i.e. line, spiral, arc). Fig.~\ref{fig_curvGraph} shows the definition of a road segment that is composed of the geometric primitive sequence line-spiral-arc-spiral-line.

As shown in Fig.~\ref{fig_curvGraph}~a), the curvature $\kappa(s)$ is a piecewise linear defined function over distance. Since we use lines, spirals and arcs, three definitions for those pieces are required. A line is described by $\kappa(s) = 0$ which means, only a length $L$ needs to be specified. An arc's course is defined by a constant curvature $\kappa(s) = \kappa_\text{arc} = \text{const}$. Therefore, an arc is defined by a length $L$ and a radius $R$. Note that we use radii in the input description because they are more intuitive than the curvature. Finally, a spiral is characterized by a linear change of curvature: $\kappa(s) = a s$, where $a$ is constant (for details on spirals cf. \cite{Baier.2008}). Consequently, in addition to the length $L$, the start and end radius $R_s$ and $R_e$ need to be defined for a spiral to be unique. Those definitions are then being concatenated to a curvature graph and result in an unambiguous road course in the $x,y$-plane as in ~\ref{fig_curvGraph}~b), if either the initial values for position and orientation at the start are provided or the road is connected to another segment without a curvature jump.

\subsection{Segments}\label{sec_segments}
In the following, the segments which might be defined in the road network input are described starting with connection roads, followed by junctions and roundabouts.

Connection roads consist of single \texttt{road} tags defined by a reference line and definition for lanes and signals in the $s,t$-coordinate system. The latter is a moving coordinate system along the distance $s$. The $t$-coordinate is defined as the offset orthogonal to the reference line at position $s$. Lane extension and merging follow 3rd-degree polynomials which are stated by their start and end point in $s,t$-coordinates, the corresponding polynomial coefficients are calculated by the road generation tool.

\begin{figure}[tb] 		
	\centering
	\includegraphics{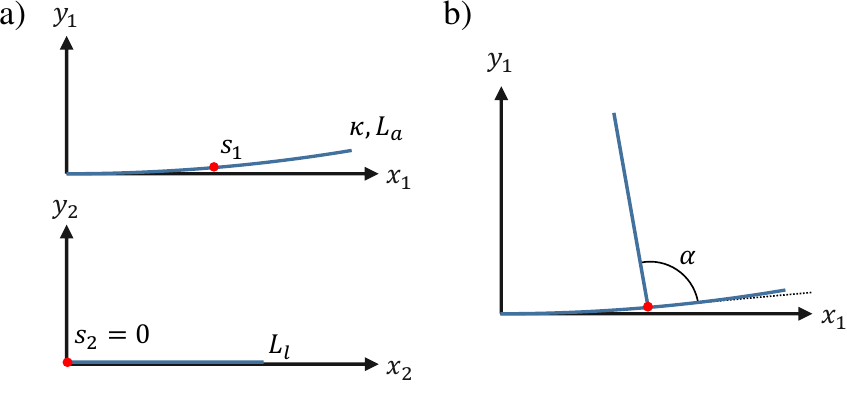}	
	\caption{Two roads are coupled to form a T-junction. Road 1 is defined as an arc by length and curvature whereas road 2 is a straight line with length $L_l$. On both roads the intersection point along $s$ is defined. In addition, the angle $\alpha$ at this point is stated. }
	\label{fig_juncDef}
\end{figure}

At least two and at most three \texttt{road} entities are required to create a T-junction. Often one road is the dominant or main road in an intersection, which can be exploited by defining it as a single road that is intersected by an access road. Alternatively, three single roads can be combined to form a T-junction by defining one reference road and the angles at which the other two roads are connected. The main idea of the intersection concept is that each road is defined in its own local coordinate system (cf. Fig.~\ref{fig_juncDef}~a)) and is combined afterwards, as shown in Fig.~\ref{fig_juncDef}~b). E.g., the T-junction in Fig.~\ref{fig_juncDef} consists of a main road described by an arc and an access road whose course follows a line. The main road is intersected by the access road approximately at its center ($s_1$) whereas the additional road is connected at its origin ($s_2=0$) under the angle $\alpha$. The reference coordinate system of the segment now is at the junction's center with the $x$-axis parallel to the tangent of the main road at this point. The described road combinations are defined within \texttt{intersectionPoint}-tags which are part of the $I$-knots of Fig.~\ref{fig_schema}.

In addition to topological information of the junction, it is possible to define the intersection area more detailed in a \texttt{coupler}-tag. First, for each road arm the distance from the intersection point to the beginning of the junction can be specified. The latter means the point where turning lanes start to leave the course of the reference line. They are specified by their minimal radius, the concrete course is calculated automatically (cf.~Sec.~\ref{sec_inters}). Fig.~\ref{fig_createJunction} shows the concept of the intersection area. Furthermore, additional lanes for left and right turns can be added ad libitum. The descriptions of T- and X-junctions are a lot alike and thus X-junctions can be created analogously to the preceding paragraph about T-junctions.

Roundabouts are defined similar to junctions with the difference that the reference road is of circular form (for now a circle) and every access road is assigned at a position $s$ with an angle $\alpha$ to this reference element.

\subsection{Combine Segments}
\begin{figure}[bt] 		
	\centering
	\includegraphics{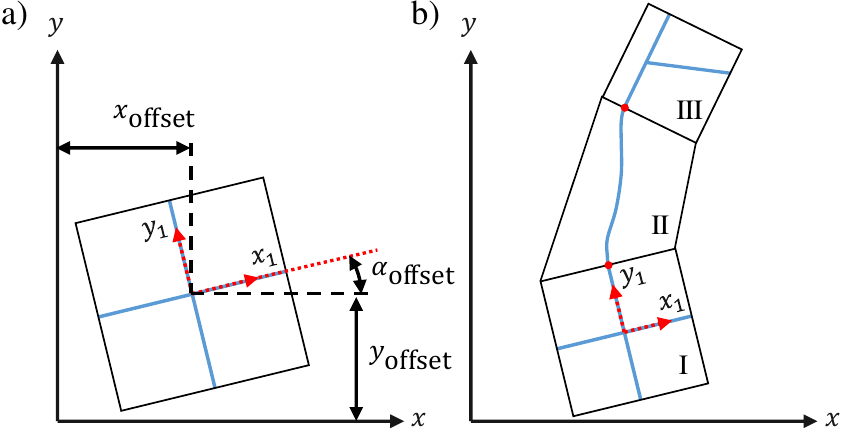}
	\caption{Three segments are concatenated to form a road network. First, one reference segment has to be set (I). For further connections only the two endpoints need to be specified because concatenation follows the continuous curvature condition.}
	\label{fig_concatSeg}
\end{figure}

Once all desired segments are specified, those may be concatenated to form a road network. In general, all segment types can be combined with each other. The only condition is that the curvature at the connection point is mutual to fulfill the continuous curvature condition we introduced for reference lines. However, in the current state of the road generation tool the curvature has to be zero at the connection to combine two segments. Fig.~\ref{fig_concatSeg} illustrates the combination of an X-junction, connection road and T-junction. In this example, segment I provides the reference coordinate system which can be placed at any position and orientation $x_\text{offset}, y_\text{offset}$ and $\alpha_\text{offset}$ in world coordinates. Next, the connections between I and II as well as II and III are defined in a linking tag by stating which segment endings are linked. With this information, the road generation tool is able to generate an OpenDRIVE map uniquely.

One more option of the road network concept is the automated generation of a compound curve (composed of two spirals with an arc between them) to close the gap between two loose road ends. This might be required if the road network should be used for endurance simulations where traffic agents shall drive in circles infinitely often and no sources and sinks are implemented in the simulation environment. It is not possible to calculate a curvature graph which matches a desired endpoint in the $x,y$-plane from an known start point analytically because of the algebraic properties of spirals (cf.~Sec.~\ref{sec_geometry}). Therefore, an optimization problem needs to be solved which attempts to find combinations of segment lengths $s_\text{i}$ and a radius $R$ which form a curve similar to Fig.~\ref{fig_curvGraph} that matches the desired end point with $\kappa=0$.

\section{Road generation} \label{sec_roadGen}
The specified input may be translated into the road network standard format OpenDRIVE by a command-line tool. Since the input describes a simplified representation of the road network, missing information needs to be calculated to provide a verified OpenDRIVE map. In the following, some steps of these calculations are described starting with details on geometric properties of roads and lanes followed by the generation of intersections. 
\subsection{Geometry}\label{sec_geometry}
	As introduced in Sec.~\ref{sec_refline}, we assume that a road's reference line is curvature continuous and therefore, a single initial position and heading is sufficient to calculate its course. However, OpenDRIVE requires a start point and heading for each line, arc or spiral. In order to provide information on $x,y$-coordinates and orientation $\varphi$ for each geometric primitive, a transformation depending on the geometry type is necessary. For a straight line, the following trivial equations result:
    \begin{equation}
        \varphi(s) = \varphi_0, \quad x(s) = s \cos(\varphi_0), \quad y(s) = s \sin(\varphi_0) \text{.}
        \label{eq_line}
    \end{equation}
    With a segment of a circle, however, the orientation also changes along $s$. With constant curvature $\kappa$ and the angle $\varphi(s) = s \kappa + \varphi_0$ this results in:
    \begin{equation}
    \begin{split}
        x(s) &= \frac{1}{\kappa} \left(\sin\varphi(s) - \sin\varphi_0 \right)\\
        y(s) &= \frac{1}{\kappa} \left(\cos\varphi_0 - \cos\varphi(s) \right) \text{.}
    \end{split}
    \label{eq_arc}
    \end{equation}
    For a spiral the coordinates along $s$ can be described with a parameter \(a = \frac{1}{\sqrt{2Rs}} = \sqrt{\frac{\kappa}{2s}}\) which is constant at any point of the spiral (cf. \cite{Baier.2008}) and the following equations hold: 
    \begin{equation} 
        x = \frac{1}{a}\int_0^{as}  \cos\left(t^2\right)dt, \quad y = \frac{1}{a}\int_0^{as}  \sin\left(t^2\right)dt \text{.}
        \label{eq_spiral}
    \end{equation}
	The formulas in (\ref{eq_spiral}) are called Fresnel integrals and it can be shown that there is no analytical solution for them. However, they can be approximated by a power series expansion for sine and cosine. With help of (\ref{eq_line}) to (\ref{eq_spiral}) missing information for each reference line can be calculated such that OpenDRIVE requirements are fulfilled.

    In a further step, lanes are added to the reference line. Each is described by a lane type, a marking, and especially by a width $w$, which for each lane $i$ corresponds to a polynomial $w_i(s)=as^3 + bs^2 + cs + d$ with coefficients \(a,b,c,d\). The width for each lane is defined separately and the final outer $t$-coordinate of the road results from the sum of all $w_i(s)$. Lanes are divided into lane sections in which those polynomials are valid. Note that the $s$ coordinate is zero at the beginning of a road but not at the beginning of a lane section. In the simplest case, $w_0$ is constant and $a=b=c=0, d=w_0$ holds. If a lane widening occurs, the lane width increases from $w_i(s_0)=0$ to the final width $w_i(s_0+ds)=w_0$ over a given length $ds$. Additional conditions for modeling the four polynomial coefficients are continuous transitions \(\left(w_i'(s) = 0\right)\) at start and end of the widening $s_0$ and $s_0 + ds$, respectively. This case, the four coefficients can be uniquely solved with help of the described conditions. A lane lapse behaves analogously but starting with $w_i(s_0)=w_0$ and ending with $w_i(s_0+ds)=0$.
        
\subsection{Intersections}\label{sec_inters}
    \begin{figure}[tb]        
        \centering
        \includegraphics{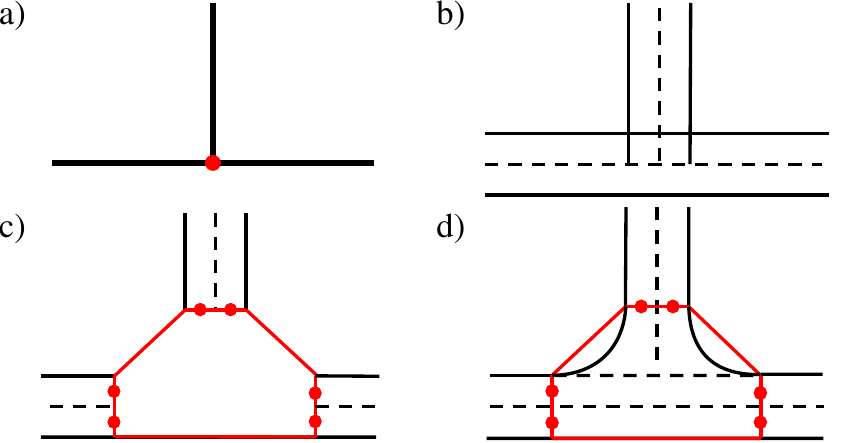}
        \caption{From logical definition to concrete intersection. After the reference lines are in the correct position a), corresponding lanes are generated (b). Subsequently, the intersection area is cut free (c) and all required connection lanes are generated automatically (d).}
        \label{fig_createJunction}
    \end{figure}
	
    Besides the reference lines, the intersections and roundabouts specified in the input file have to be processed in order to form valid OpenDRIVE entities. Each intersection or roundabout consists, as shown in Fig.~\ref{fig_schema}, of corresponding roads \(r_i\) as well as more detailed information \(I\). Besides the size of the intersection area, it is also specified which lanes of the individual roads are connected or whether additional lanes are added before the intersection.

    A segment is constructed such that the junction midpoint lies in the origin and intersects the reference road without any angle. Further roads \(r_i\) intersect the origin at the specified road position \(s_\text{I,i}\) at a defined angle \(\alpha_\text{I,i}\), so that the starting point and the starting angle of each geometry of this road must be shifted by a translatory offset \(\tilde{x}, \tilde{y}\) and rotated around the angle \(\tilde{\varphi} = \alpha_\text{I,i}\). The translatory offset can be determined from the coordinates at position \(s_\text{I,i}\) so that all segments with their starting points \(x_0,y_0,\varphi_0\) can be transformed as follows.

    \begin{align} \label{eq_transform}
        \begin{split}
            x &= \cos(\tilde{\varphi}) \left(x_0-\tilde{x}\right) - \sin(\tilde{\varphi}) \left(y_0-\tilde{y}\right) \\
            y &= \sin(\tilde{\varphi}) \left(x_0-\tilde{x}\right) + \cos(\tilde{\varphi}) \left(y_0-\tilde{y}\right) \\
            \varphi &= \varphi_0 - \tilde{\varphi} \text{.} 
        \end{split}
    \end{align}

    Before inserting new connecting roads, the junction area has to be generated, which requires a cut in the reference line of each road $r_i$. From the junction midpoint at \(s_\text{I,i}\) we define \(s_A\) as the length along $s$ which spans the junction area in one direction. Note that for simplicity we assume \(s_A\) in both directions of the reference line in the documentation. In practice those are two variables. From the defined reference line of $r_\text{i}$, two parts are extracted: the part up to the beginning of the junction \((-\inf,s_I - s_A]\) and the part behind the junction \([s_I + s_A, \inf)\). The two relevant parts of the reference line are now iteratively copied over all geometries to generate the new exit roads of the intersection. Depending on the cut of a straight line, arc or spiral, the respective segment of the reference line has to be adapted. The result of this step can be illustrated as shown in Fig.~\ref{fig_createJunction}~c).
    


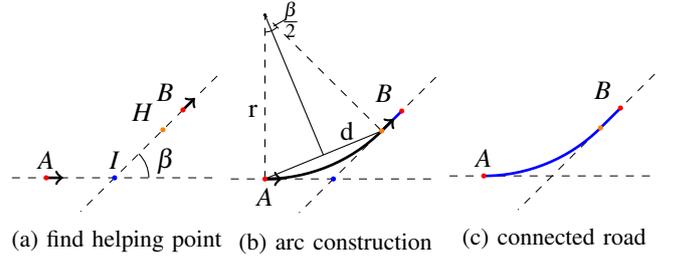
\begin{figure}
    	\begin{subfigure}{0.32\columnwidth}
        \vspace{1.05cm}
        \begin{tikzpicture}[scale=0.45]
            \coordinate[label=above:$A$] (A) at (0,0);
            \coordinate[label=135:$B$] (B) at (4,2);
            \coordinate[label=above:$I$] (iP) at (2,0);
            \coordinate[label=135:$H$] (H) at (3.41421,1.41421);
        
            \draw[dashed, line width=0.2] (-1,0) -- (5,0);
            \draw[dashed, line width=0.2] (5,3) -- (1,-1);
            \draw (3,0) arc (0:45:1);
        
            \draw[->, line width=1] (A) -- (0.5,0);
            \draw[->, line width=1] (B) -- (4.35,2.35);
            \fill[red] (A) circle (0.075);
            \fill[red] (B) circle (0.075);
            \fill[blue] (iP) circle (0.075);
            \fill[orange] (H) circle (0.075);
        
            \draw (3.5, 0.5) node {$\beta$};
        \end{tikzpicture}
        \caption{find helping point}
        \label{fig_construct_hp}
        \end{subfigure}
        \begin{subfigure}{0.32\columnwidth}
            \begin{tikzpicture}[scale=0.45]
                \coordinate[label=below:$A$] (A) at (0,0);
                \coordinate[label=135:$B$] (B) at (4,2);
                \coordinate[] (iP) at (2,0);
                \coordinate[] (H) at (3.41421,1.41421);
                \coordinate[] (E) at (0,4.8284);
                \coordinate[] (F) at ($(A)!0.5!(H)$);
            
                \draw[dashed, line width=0.2] (-1,0) -- (5,0);
                \draw[dashed, line width=0.2] (5,3) -- (1,-1);
            
                \draw[blue, line width = 1] (H) -- (B);
            
                \draw[->, black, line width=1] (A) -- (0.5,0);
                \draw[->, black, line width=1] (H) -- (3.7676,1.7676);
            
                \draw[label=below:$d$] (A) -- (H);
                \draw[line width=1] (A) arc (-90:-45:4.8284);
                \draw (0,4.3284) arc (-90:-45:0.5);
                \draw[dashed] (A) -- (E) -- (H);
                \draw[] (E) -- (F);
            
                \fill[red] (A) circle (0.075);
                \fill[red] (B) circle (0.075);
                \fill[blue] (iP) circle (0.075);
                \fill[orange] (H) circle (0.075);
                \fill[black] (E) circle (0.05);
            
                \draw (-0.35,2)  node {r};
                \draw (2.4,1.4) node {d};
                \draw (0.25,4.5) -- (0.6,4.7);
                \draw (0.75, 4.7) node {$\frac{\beta}{2}$};
            
            \end{tikzpicture}
        \caption{arc construction}
        \end{subfigure}
        \begin{subfigure}{0.32\columnwidth}
            \vspace{1cm}
        \begin{tikzpicture}[scale=0.45]
            
            \coordinate[label=above:$A$] (A) at (0,0);
            \coordinate[label=135:$B$] (B) at (4,2);
            \coordinate[] (H) at (3.41421,1.41421);
        
            \draw[dashed, line width=0.2] (-1,0) -- (5,0);
            \draw[dashed, line width=0.2] (5,3) -- (1,-1);
        
            \draw[blue, line width=1] (H) -- (B);
            \draw[blue, line width=1] (A) arc (-90:-45:4.8284);
        
            \fill[red] (A) circle (0.075);
            \fill[red] (B) circle (0.075);
            \fill[orange] (H) circle (0.075);
        
        \end{tikzpicture}
        \caption{connected road}
        \end{subfigure}
    	\caption{Process for connecting roads in junction area.}
    	\label{fig_connecting_road}
    \end{figure}

    After every street is cut and correctly positioned, the next step is calculating the reference lines of the lanes inside of the intersection area. Each lane is generated as an individual street with its reference line, whereby the single-lane lies to the right of the reference line. Based on starting point \(A\) and endpoint \(B\) as well as the angles \(\varphi_{A}\) and \(\varphi_{B}\) the new reference line of a lane is calculated. As a simplification, only straight lines and arcs are used since the discontinuity in the curvature can be neglected at low speeds. Note that the assumption still yields continuous angles along $s$. The setup for one connection road is illustrated in Fig.~\ref{fig_connecting_road}. $A$ is the end of the road entering the intersection area and $B$ is the corresponding start of the newly generated road leaving the junction. In the following, the geometric construction of the new reference line is described.

    In the first step, the endpoints are extended with straight lines to calculate the intersecting point \(I\). 
    \begin{align}
        \begin{split}
            I_x &= \frac{B_y - A_y + \left(\tan(\varphi_A) A_x - \tan(\varphi_B) B_x\right)}{\tan(\varphi_A) - \tan(\varphi_B)} \\
            I_y &= \tan(\varphi_A) \cdot I_x + \left(A_y - \tan(\varphi_A) A_x\right) \text{.}
        \end{split}     
    \end{align}
    
\begin{figure*}
	\includegraphics[width=\textwidth]{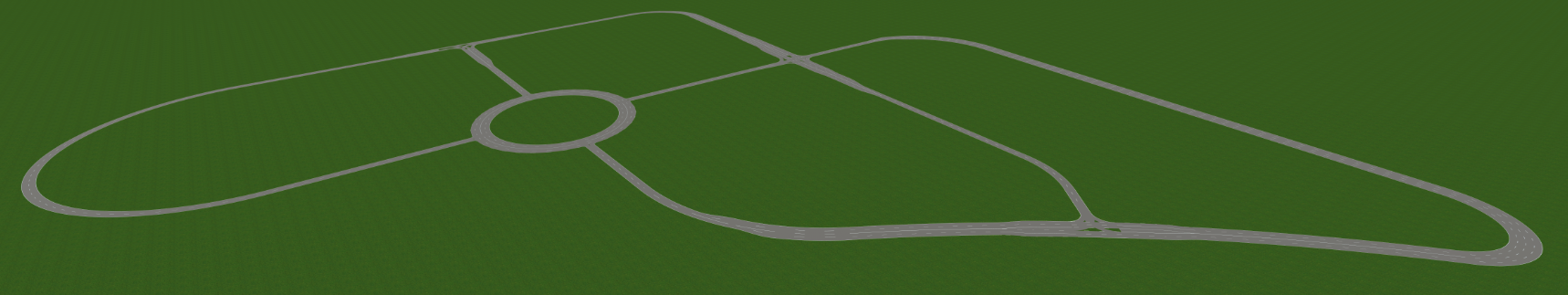}
	\caption[dummy2]{OpenDRIVE output of a road network generated by the tool which processes the developed logical road description. The road network consists of three junctions, one roundabout, one connection road and four automatically generated road network closing roads. Visualization with internal Unity\footnotemark~tool.}
	\label{fig_resHuge}
\end{figure*} 

    The euclidean distances $d_\text{A} = \|A-I\|$ and $d_\text{B} = \|B-I\|$ between endpoints and intersection point are then being calculated and the road with greater distance is extended by a line until the new distances are equal, which introduces the new helping point \(H\) at the end of the extended line. Finally, the two points are connected by an arc with radius \(r\) and length \(l\). The following equation determines the parameters of the circle, where \(\beta = \varphi_B - \varphi_A\):
    \begin{align}
        \begin{split}           
        r &= \frac{\sqrt{\left(H_x - A_x\right)^2 + \left(H_y - A_y\right)^2}}{2 \cdot \sin\left(\frac{\beta}{2}\right)} \\
        l &= \vert r \cdot \left(\varphi_B - \varphi_A\right) \vert \text{.}  
        \end{split}
    \end{align}
    
    This procedure is repeated until all connection roads, i.e. connection lanes, inside the junction are calculated.
   
\section{Results} \label{sec_res}
\footnotetext{https://unity.com/}
Recall Fig.~\ref{fig_motivation}, which shows a generated X-junction. The input description of this map consists of $32$ lines whereas the resulting OpenDRIVE file is $870$ lines long. This shows one benefit of the proposed road description which is straightforward to maintain.

\begin{figure}
	\includegraphics[width=\columnwidth]{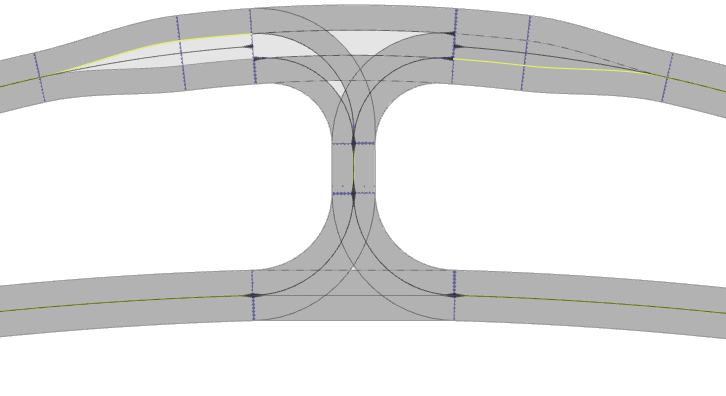}
	\caption{OpenDRIVE Output of two concatenated T-junctions. Both junctions are almost identically defined. The only differences for the upper junction are that the radius of the main road is changed and an option that automatically generates lane widening for left turns is activated. Note that black lines indicate lane courses, not their marking. Visualization in CarMaker 8.0.}
	\label{fig_res3_2}
\end{figure}

In Fig.~\ref{fig_res3_2} the OpenDRIVE output of two concatenated T-junctions is shown. Each segment consists of a main road which follows an arc and one access road that is modeled as a straight line. The angle between both roads is $\alpha=\frac{\pi}{2}$. Further, a link between both junctions is applied at the end of the respective access road. Both junctions are defined mostly the same, the only differences for the upper junction are that the radius of the main road is smaller and that an option is activated which automatically generates lane widenings for left turns on main roads. In fact, only two chars are different in the respective junction definitions which makes the description easy to vary. In addition, the input file consists of four defined roads whereas the resulting OpenDRIVE file contains 18 road definitions. These effects have impact on the overall amount of code which can be seen in Table~\ref{tab_comparison} line~2. The input file requires only $5.3\%$ as many chars as the concrete OpenDRIVE file.

\begin{table}[h]
	\caption{Comparison of OpenDRIVE and logical description in terms of amount of code.}
	\label{tab_comparison}
	\def\arraystretch{1.5}
	\begin{center}
		\begin{tabular}{l|cccc}
			\multirow{2}{*}{File} & \multicolumn{2}{c}{OpenDRIVE} & \multicolumn{2}{c}{Logical Description}\\
			& Lines & Chars & Lines & Chars \\
			\hline
			XJunc (Fig.~\ref{fig_motivation}) & $870$ & $32,468$& $3.7\%$ &$4.1\%$\\
			TwoTJunc (Fig.~\ref{fig_res3_2}) & $937$ & $36,414$& $5.0\%$ &$5.3\%$\\
			HugeRoadNet (Fig.~\ref{fig_resHuge}) & $5,477$ & $210,215$& $4.0\%$ &$4.8\%$
		\end{tabular}
	\end{center}
\end{table}

\addtolength{\textheight}{-10.8cm}   
As a general overview, in Fig.~\ref{fig_resHuge} the result for a considerably large road network is illustrated. It is described entirely by the presented logical description and then is instantiated as OpenDRIVE with help of the developed tool. The network consists of three intersections, one roundabout, one connection road and four instances of the close road network function which generates compound curves between segments automatically. As shown in Table~\ref{tab_comparison} line~3, the input file is significantly smaller in terms of code length than the resulting OpenDRIVE file. In addition, only little parts of the input have to be changed (e.g. the curvature of one connection road) in order to obtain a road network of similar topology but with thousands of changes in the OpenDRIVE file. This way, parameter spaces for distinct quantities may be explored and many concrete road networks can be generated using one \emph{logical} definition.

All generated OpenDRIVE files are verified against current OpenDRIVE XSD schemas of version 1.4\footnote{http://www.opendrive.org/tools/OpenDRIVE\_1.4H.xsd} and 1.5\footnote{http://www.opendrive.org/tools/OpenDRIVE\_1.5M.xsd}. In addition, we evaluated selected tracks with ODR Viewer\footnote{http://www.opendrive.org/tools/odrViewer.zip}, CarMaker 8.0 and RoadRunner\footnote{https://www.vectorzero.io/roadrunner} by making sure the import works, the visualization is plausible and a vehicle is able to pass the generated track.

\section{Conclusion and outlook}\label{sec_concl}
In this paper, we introduced a prototypical concept to describe road networks in a logical manner with the possibility to generate a standardized concrete format which can be used in most simulation environments.
The results are ready to use simulation maps which can be easily varied.

A next step could be the analysis of real road networks to calibrate the assumed parameters, e.g. for main and access roads or radii and width of lanes inside junctions.
Further, the logical input and road generation tool might be extended with more features such as nested junctions or objects next to the road.
In addition, the concatenation of segments could be improved. Curvatures unequal to zero might be allowed at the connection between two segments, or the close road network function could be enhanced such that linking elements among two junctions can be generated automatically.





\bibliographystyle{IEEEtran}
\bibliography{IEEEexample,ma_russ,anonymization,roadgen}

\end{document}